\documentclass[epj]{webofc}
\usepackage[varg]{txfonts}   
\usepackage{bm}

\newcommand{\beq}{\begin{equation}}
\newcommand{\eeq}{\end{equation}}
\newcommand{\beqy}{\begin{eqnarray}}
\newcommand{\eeqy}{\end{eqnarray}}

\newcommand{\me}[3]{\langle #1\vert\ #2\ \vert #3\rangle}

\woctitle{(MENU 2013)}
\begin{document}
\title{Excited states in lattice QCD with the stochastic LapH method}
\author{
     John Bulava\inst{1}
\and Brendan Fahy\inst{2}
\and Justin Foley\inst{3}
\and You-Cyuan Jhang\inst{2}
\and Keisuke J. Juge\inst{4,5}
\and David Lenkner\inst{2}
\and Colin Morningstar\inst{2}\fnsep\thanks{Speaker.}
\and Chik Him Wong\inst{6} 
}

\institute{
 School of Mathematics, 
 Trinity College, Dublin 2, Ireland 
\and
 Dept.~of Physics, Carnegie Mellon University, 
 Pittsburgh, PA 15213, USA
\and
 Dept.~of Physics and Astronomy, University of Utah, 
 Salt Lake City, UT 84112, USA
\and
 High Energy Accelerator Research Organization (KEK), 
 Ibaraki 305-0801, Japan
\and
 Dept.~of Physics, University of the Pacific, 
 Stockton, CA 95211, USA
\and
 Dept.~of Physics, University of California San Diego,
 La Jolla, CA 92093, USA
}

\abstract{
Progress in computing the spectrum of excited baryons and mesons 
in lattice QCD is described.  Results in the zero-momentum 
bosonic $I=\frac{1}{2},\ S=1,\ T_{1u}$ 
symmetry sector of QCD using a correlation matrix of 58 operators are presented.
All needed Wick contractions are efficiently evaluated using a stochastic method 
of treating the low-lying modes of quark propagation that exploits Laplacian 
Heaviside quark-field smearing.  Level identification using
probe operators is discussed.
}
\maketitle
\section{Introduction}

In a series of papers\cite{baryons2005A,baryon2007,nucleon2009,
Bulava:2010yg,StochasticLaph,ExtendedHadrons,lattice2013}, we have been striving to 
compute the finite-volume stationary-state energies of QCD using Markov-chain
Monte Carlo integration of the QCD path integrals formulated on a
space-time lattice.  In this talk, our progress towards this goal is described.  
First results in the zero-momentum bosonic $I=1,\ S=0,\ T_{1u}^+$ symmetry sector 
of QCD using a correlation matrix of 56 operators were recently presented
in Ref.~\cite{lattice2013}.  Here, preliminary results in the zero-momentum 
bosonic $I=\frac{1}{2},\ S=1,\ T_{1u}$ sector using a correlation matrix of 
58 operators are presented.  Nine spatially-extended single-kaon operators are
used, and 49 two-meson operators involving a wide variety of light isovector, 
isoscalar, and strange meson operators of varying relative momenta are
included.  All needed Wick contractions are efficiently evaluated using a 
stochastic method of treating the low-lying modes of quark propagation that 
exploits Laplacian Heaviside quark-field smearing.   Given the large number of 
levels extracted, level identification becomes a key issue.

\section{Energies from correlations of single-meson and two-meson operators}
\label{sec:ops}

The stationary-state energies in a particular symmetry sector can be extracted 
from an $N\times N$ Hermitian correlation matrix 
   $ {\cal C}_{ij}(t)
   = \langle 0\vert\, O_i(t\!+\!t_0)\, \overline{O}_j(t_0)\ \vert 0\rangle,
   $
where the $N$ operators $\overline{O}_j$ act on the vacuum to create the states 
of interest at source time $t_0$ and are accompanied by conjugate operators $O_i$ 
that can annihilate these states at a later time $t+t_0$.  
Estimates of ${\cal C}_{ij}(t)$ are obtained with the Monte Carlo method
using the stochastic LapH method\cite{StochasticLaph} which allows all needed
quark-line diagrams to be computed.  The operators that we use have been
described in detail in Refs.~\cite{baryons2005A,StochasticLaph,ExtendedHadrons}.
All of our single-hadron operators are assemblages of basic building blocks
which are gauge-covariantly-displaced, LapH-smeared quark fields.
We simplify our spectrum calculations as much as possible by working with 
single-hadron operators that transform irreducibly under all symmetries of a 
three-dimensional cubic lattice of infinite extent or finite extent with 
periodic boundary conditions. We construct our two-hadron operators as 
superpositions of single-hadron operators of definite momenta.
The details are described in Ref.~\cite{ExtendedHadrons}.
This approach is efficient for creating large numbers of two-hadron 
operators, and generalizes to three or more hadrons.
We utilize multi-hadron operators with a variety of different relative momenta.

\begin{figure}[t]
\begin{center}
\includegraphics[width=1.40in]{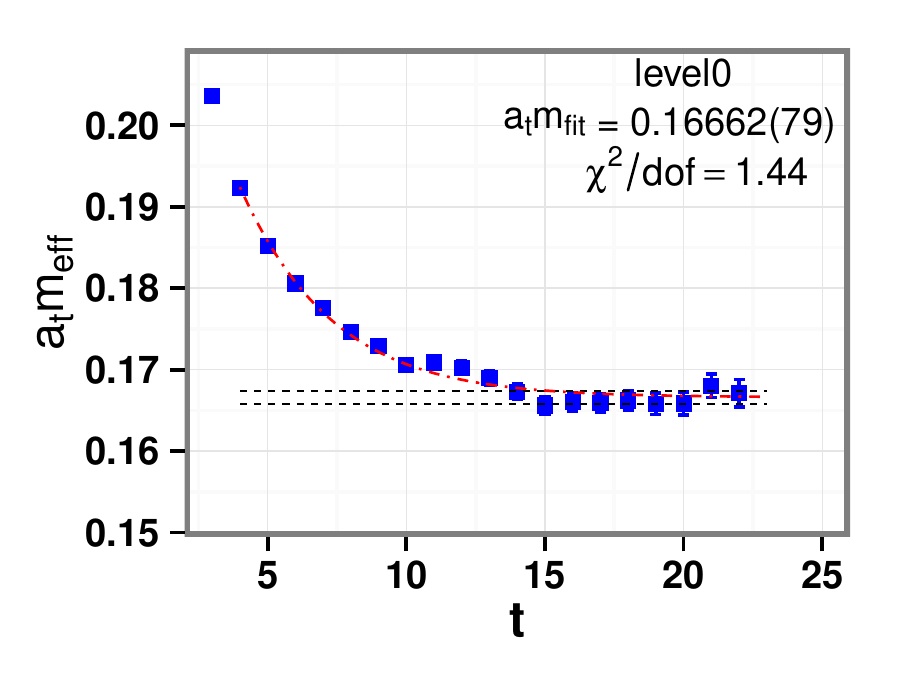}%
\includegraphics[width=1.40in]{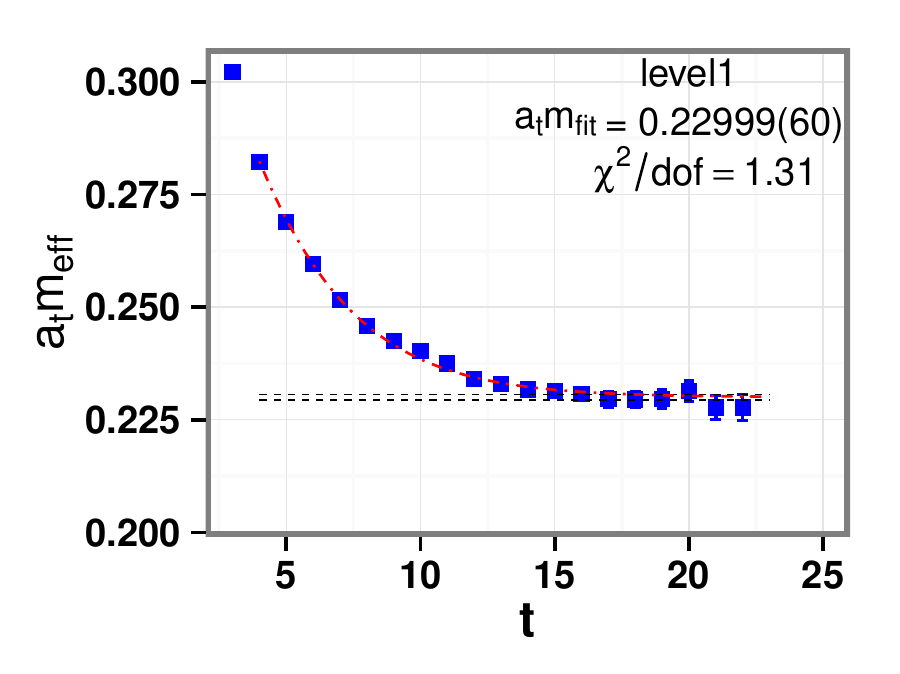}%
\includegraphics[width=1.40in]{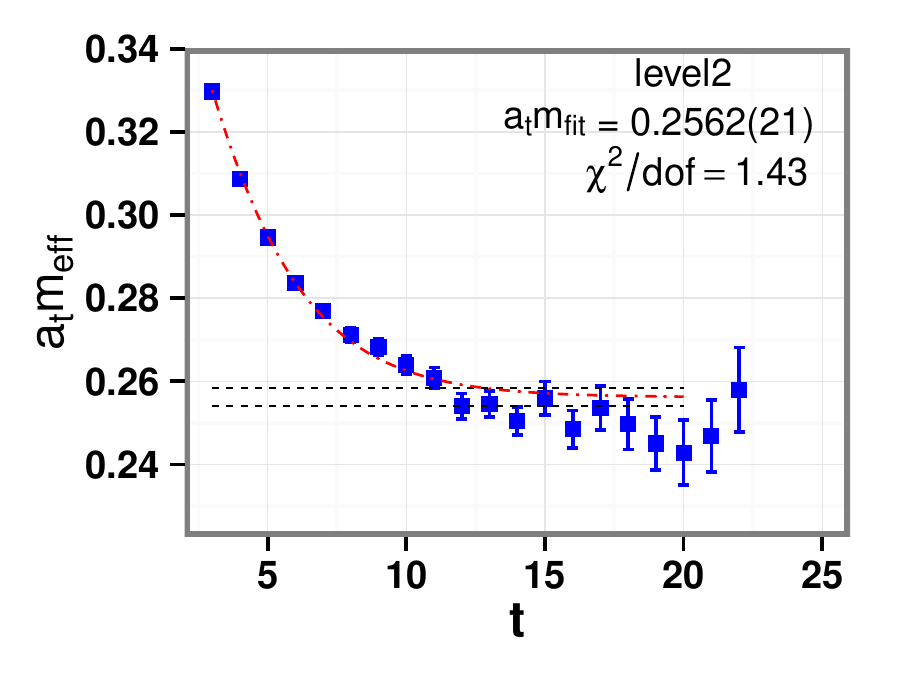}%
\includegraphics[width=1.40in]{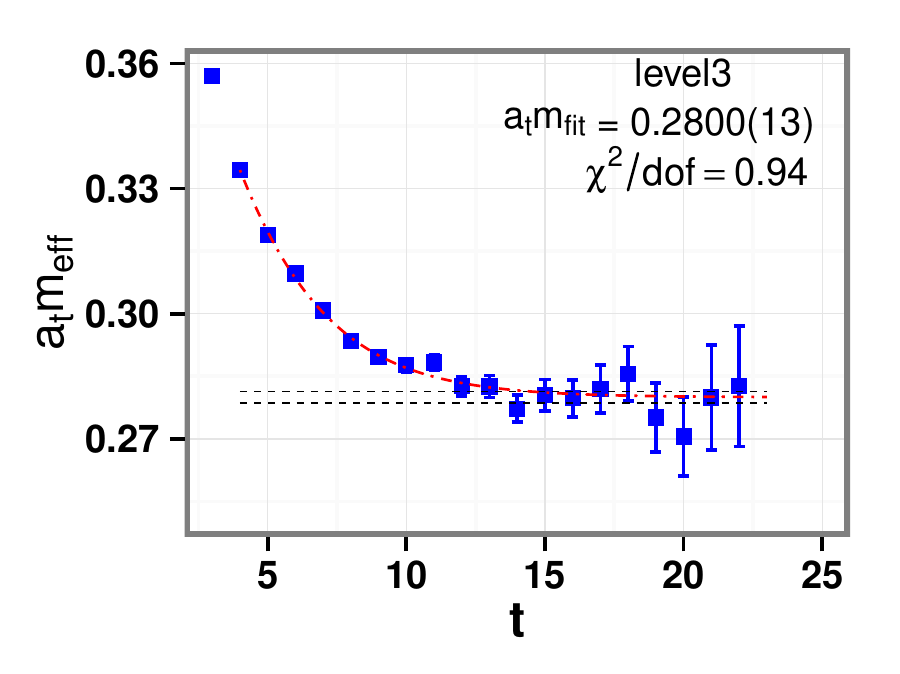}%
\\
\includegraphics[width=1.40in]{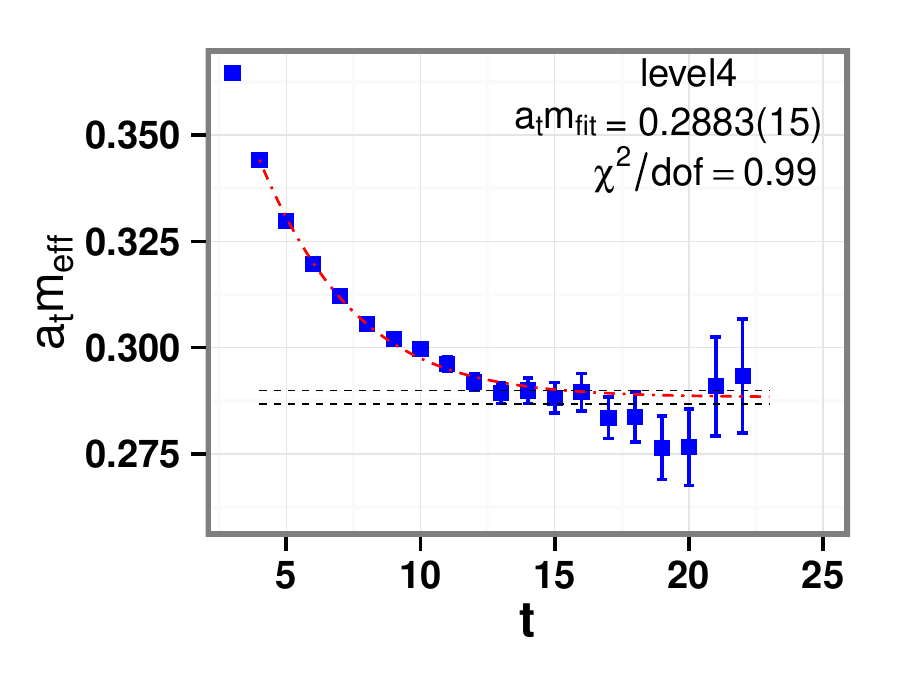}%
\includegraphics[width=1.40in]{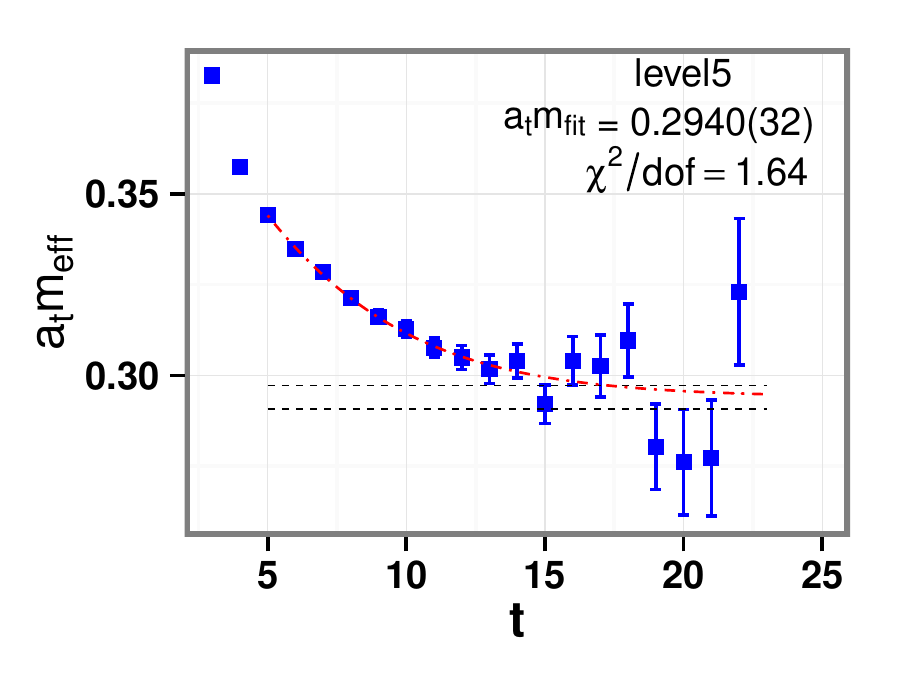}%
\includegraphics[width=1.40in]{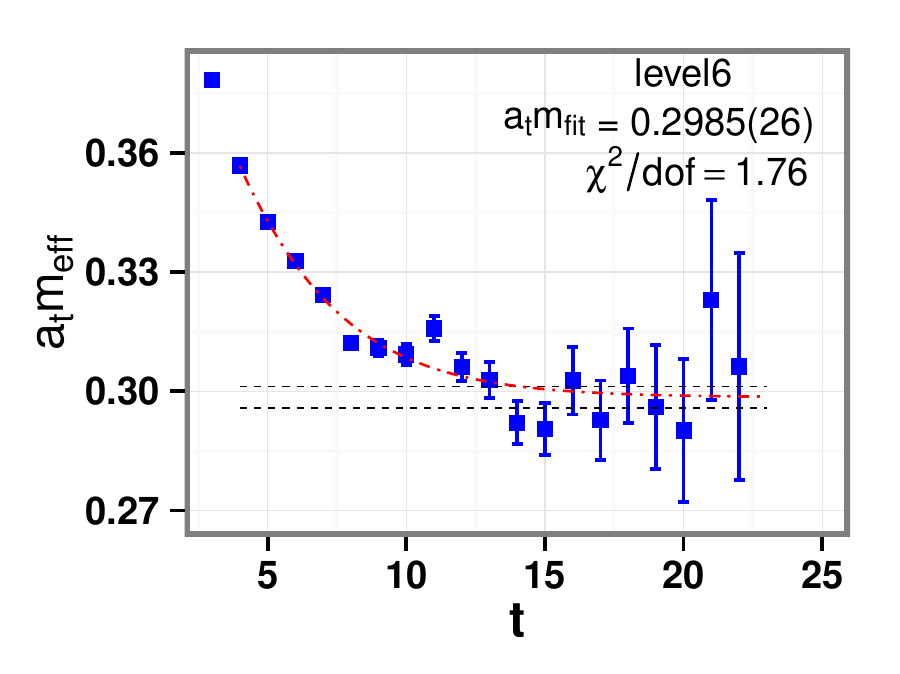}%
\includegraphics[width=1.40in]{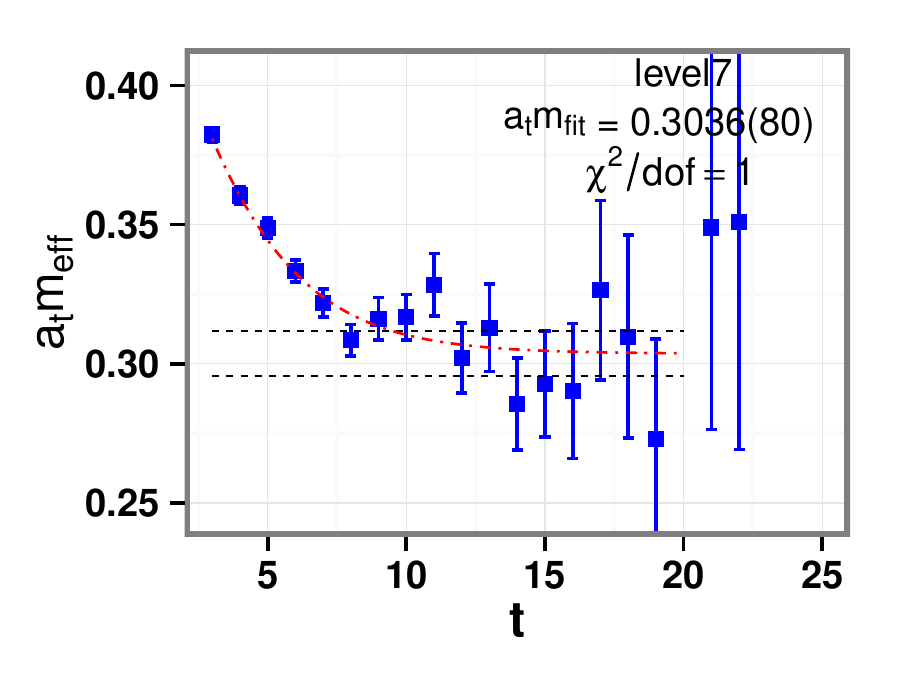}%
\\
\includegraphics[width=1.40in]{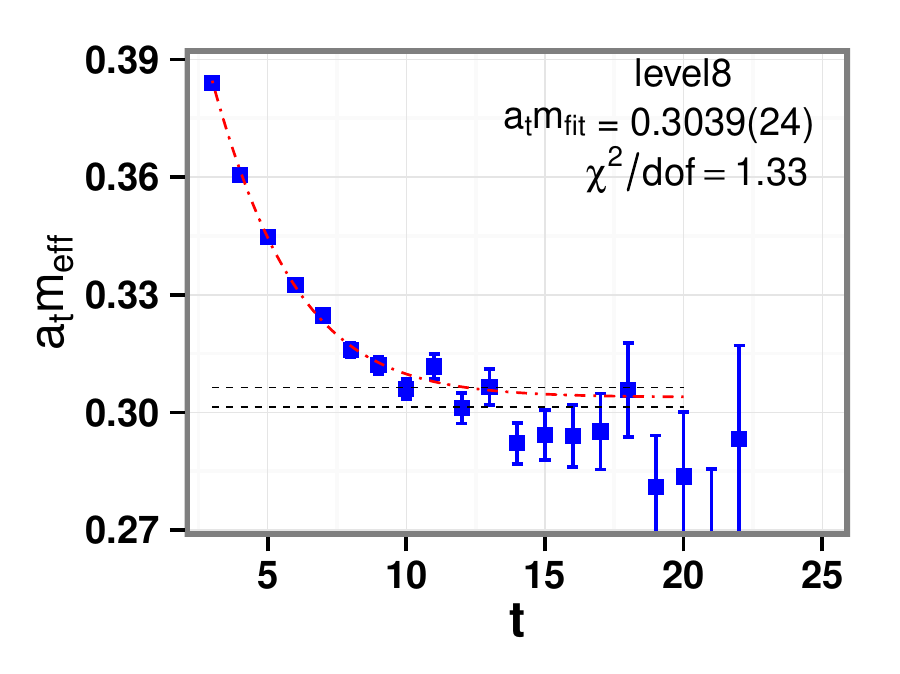}%
\includegraphics[width=1.40in]{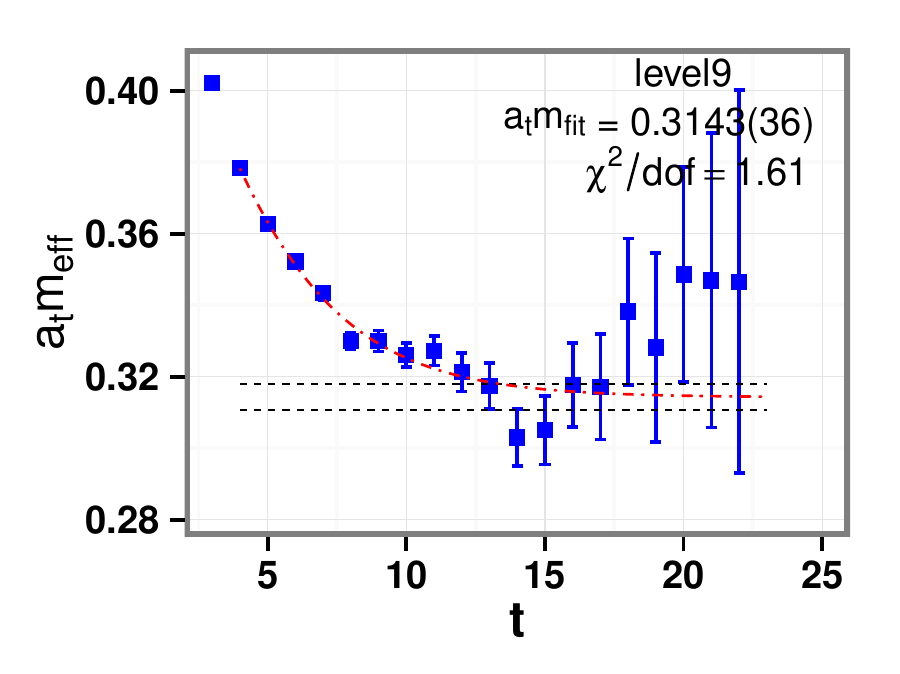}%
\includegraphics[width=1.40in]{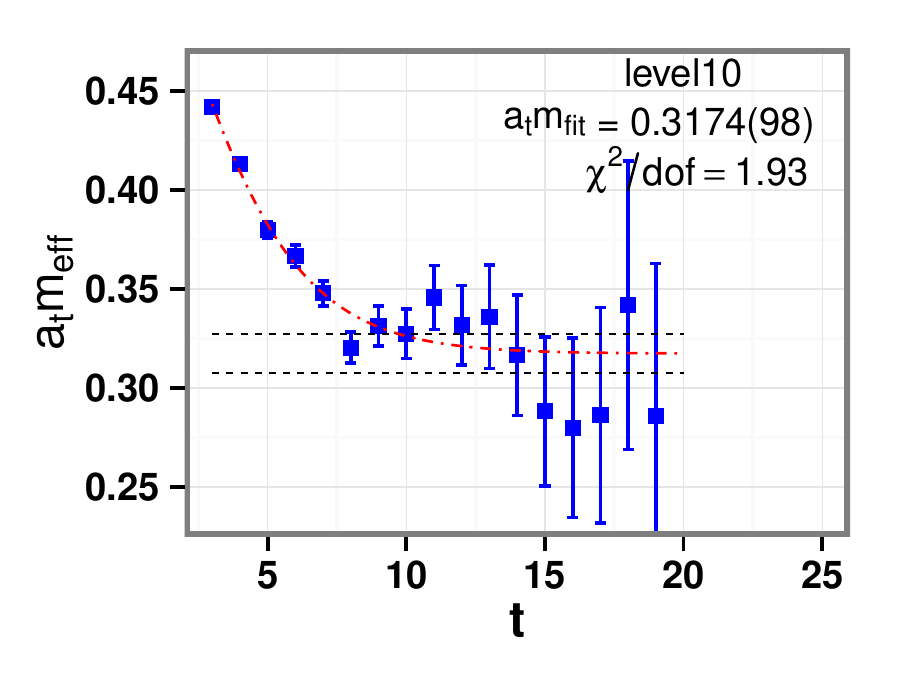}%
\includegraphics[width=1.40in]{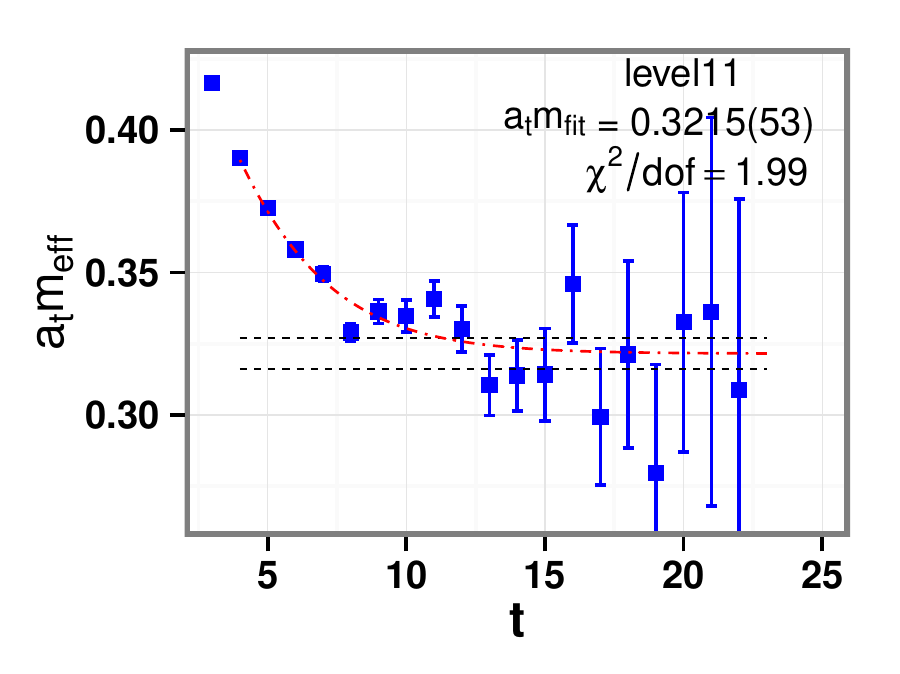}%
\end{center}
\vspace*{-7mm}
\caption[cap1]{
Rotated effective masses $m_G^{(n)}(t)$ 
(see Eq.~\ref{eq:roteffmass}) for the 12 lowest-lying energy 
levels in the zero-momentum bosonic $I=\frac{1}{2},\ S=1,\ T_{1u}$
channel for the $(24^3\vert 390)$ ensemble using 9 single-meson
operators, 25 kaon+isovector operators, 
12 kaon+$(\overline{u}u\!+\!\overline{d}d)$-isoscalar operators, and 
12 kaon+$\overline{s}s$-isoscalar operators.  The two horizontal dashed lines
in each plot indicate the best-fit energy from the correlated-$\chi^2$ fits.
The red dashed-dotted line in each plot shows the best-fit two-exponential
function.  Fit results and qualities are also listed in each plot.
Higher-lying levels cannot be shown here due to page limitations.
Over 50 energies were extracted.
\label{fig:levels1}}
\end{figure}  

In finite volume, all energies are discrete so that each correlator matrix
element has a spectral representation of the form
$   C_{ij}(t) = \sum_n Z_i^{(n)} Z_j^{(n)\ast}\ e^{-E_n t},$
with $Z_j^{(n)}=  \me{0}{O_j}{n}$,
assuming temporal wrap-around (thermal) effects are negligible.
It is not practical to extract the $E_n$ and $Z_j^{(n)}$ from fits
to all of our correlator matrix elements.  Instead, we rotate the matrix 
so that its off-diagonal elements are statistically consistent with zero 
for large time separations.  The rotated correlator is given by
$
 G(t) = U^\dagger\ C(\tau_0)^{-1/2}\ C(t)\ C(\tau_0)^{-1/2}\ U,
$
where the columns of $U$ are the orthonormalized eigenvectors of
$C(\tau_0)^{-1/2}\ C(\tau_D)\ C(\tau_0)^{-1/2}$ for a judicious choice
of $\tau_0$ and $\tau_D$.
Rotated effective masses can then be defined by
\beq
  m_G^{(n)}(t)=\frac{1}{\Delta t}
  \ln\left(\frac{G_{nn}(t)}{
  G_{nn}(t+\Delta t)}\right),
\label{eq:roteffmass}
\eeq
using $\Delta t=3$.  These tend to the lowest-lying $N$ stationary-state energies
produced by the $N$ operators.  Correlated-$\chi^2$ fits to 
the estimates of $G_{nn}(t)$ using the forms 
$A_n e^{-E_n\,t}(1+B_n e^{-\Delta_n^2\,t})$ yield the energies $E_n$
and the overlaps $A_n$ to the rotated operators for each $n$. 

We are currently focusing on three Monte Carlo ensembles: (A) a set of 
412 gauge-field configurations on a large $32^3\times 256$ anisotropic lattice 
with a pion mass $m_\pi\sim 240$~MeV, (B) an ensemble of 551 configurations
on an $24^3\times 128$ anisotropic lattice with a pion mass
$m_\pi\sim 390$~MeV, and (C) an ensemble of 584 configurations
on an $24^3\times 128$ anisotropic lattice with a pion mass
$m_\pi\sim 240$~MeV.  We refer to these ensembles as the 
$(32^3\vert 240)$, $(24^3\vert 390)$, and $(24^3\vert 240)$ ensembles,
respectively.  

Here, we focus on the resonance-rich
$I=\frac{1}{2},\ S=1,\ T_{1u}$ channel of total zero momentum.  This channel
has odd parity, and contains the spin-1 and spin-3 mesons.
A partial sampling of our ``first-pass'' results for the $(24^3\vert 390)$ 
ensemble obtained from a $58\times 58$ correlation matrix
is presented in Fig.~\ref{fig:levels1}.
The results shown here are not finalized yet.  We are still
varying the fitting ranges to improve the $\chi^2$, as needed in
some instances.  We are investigating the effects of adding more
operators, and we are even still verifying our analysis/fitting
software.  However, these figures do demonstrate that the extraction
of a large number of energy levels is indeed possible, and the
plots indicate the level of precision that can be attained with
our stochastic LapH method.  Keep in mind that we have not included 
any three-meson operators in our correlation matrix.

\begin{figure}[t]
\begin{center}
\includegraphics[width=1.40in]{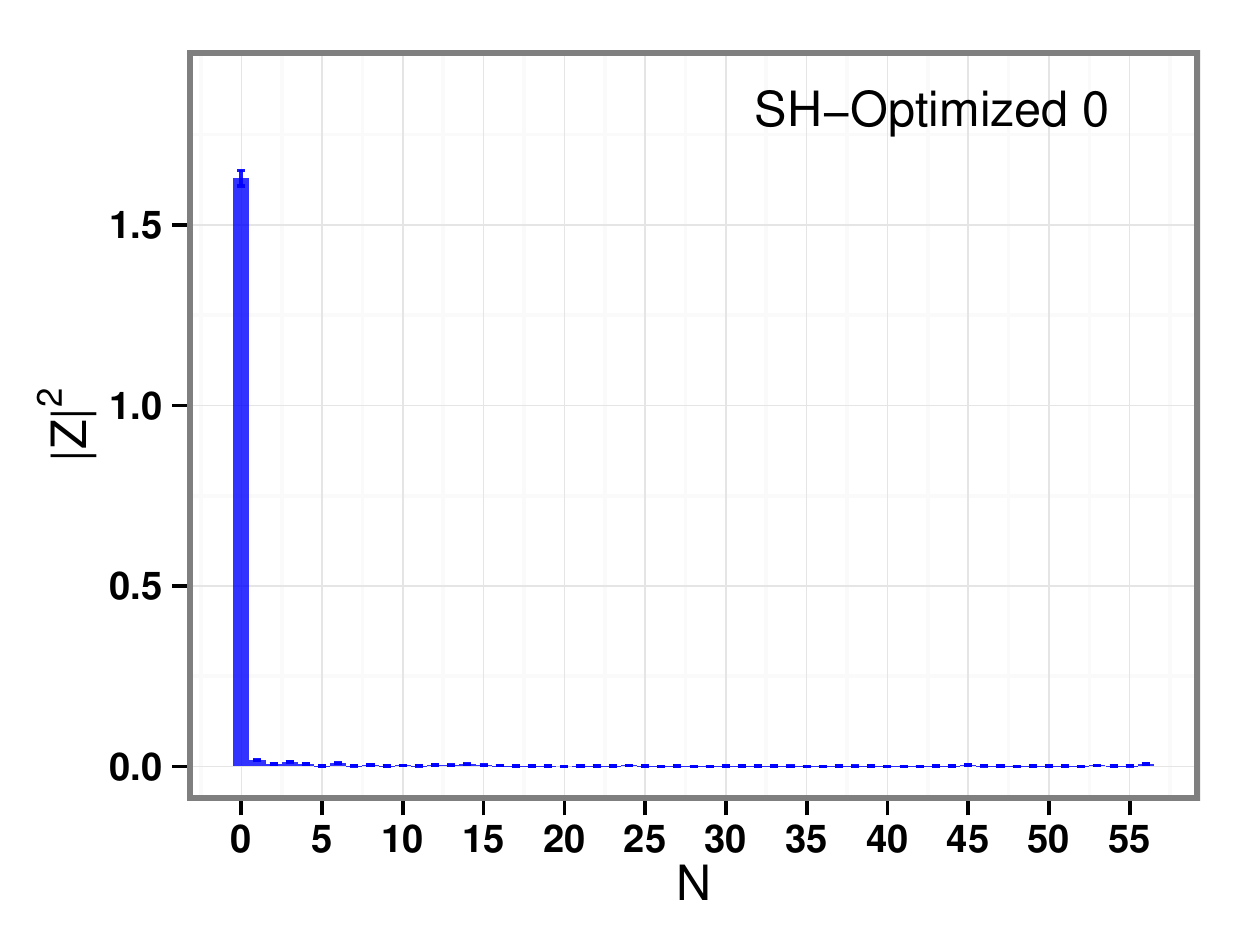}%
\includegraphics[width=1.40in]{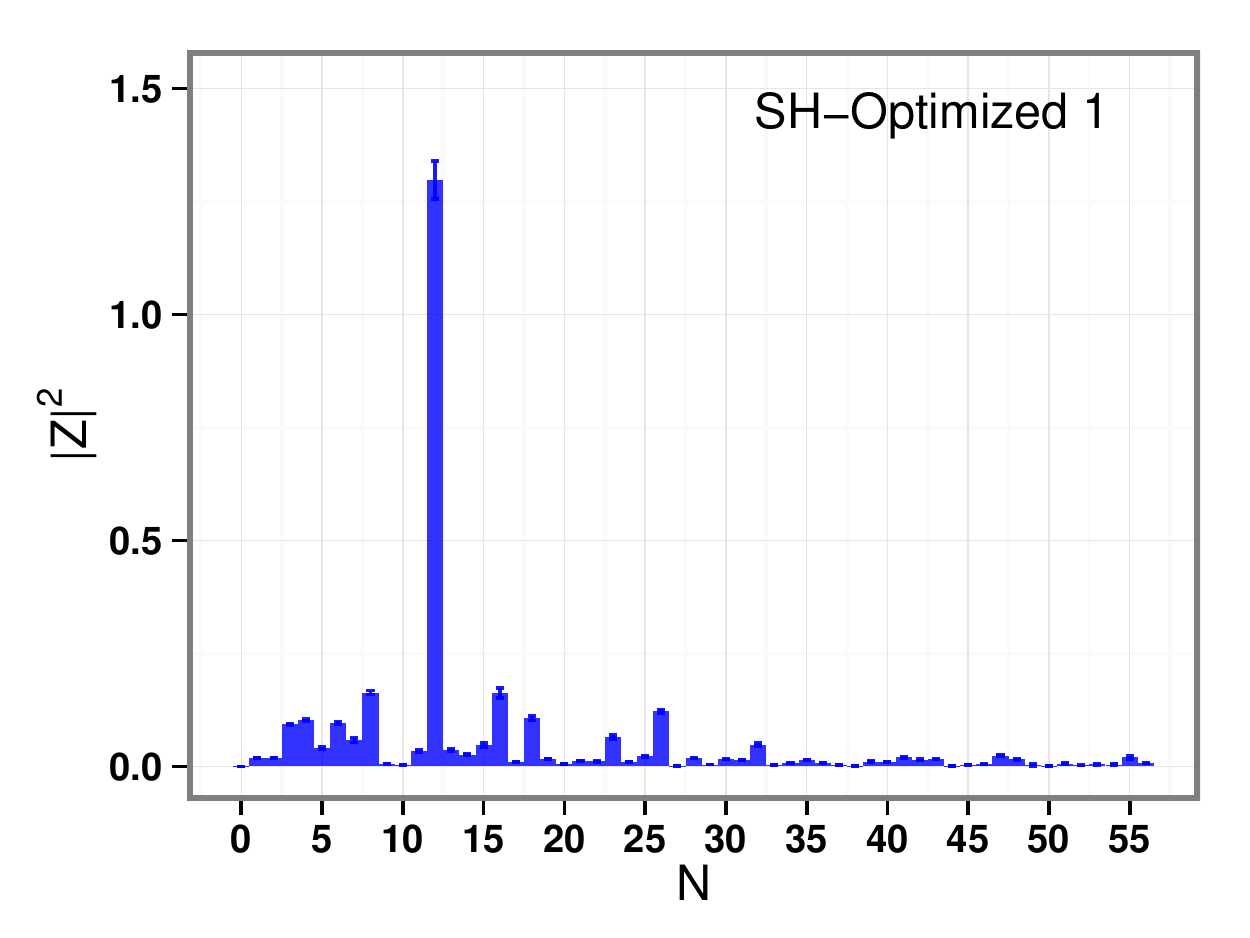}%
\includegraphics[width=1.40in]{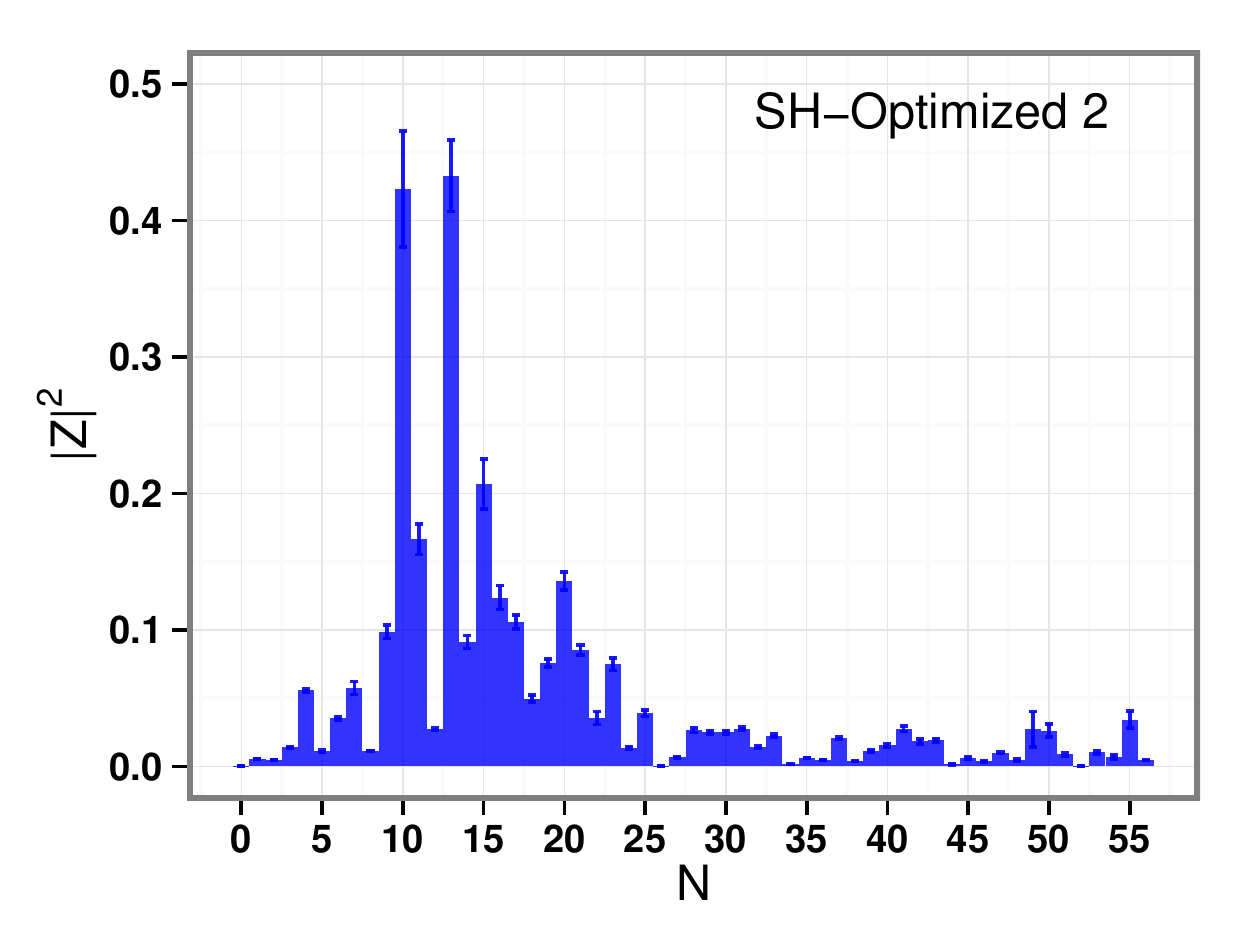}%
\includegraphics[width=1.40in]{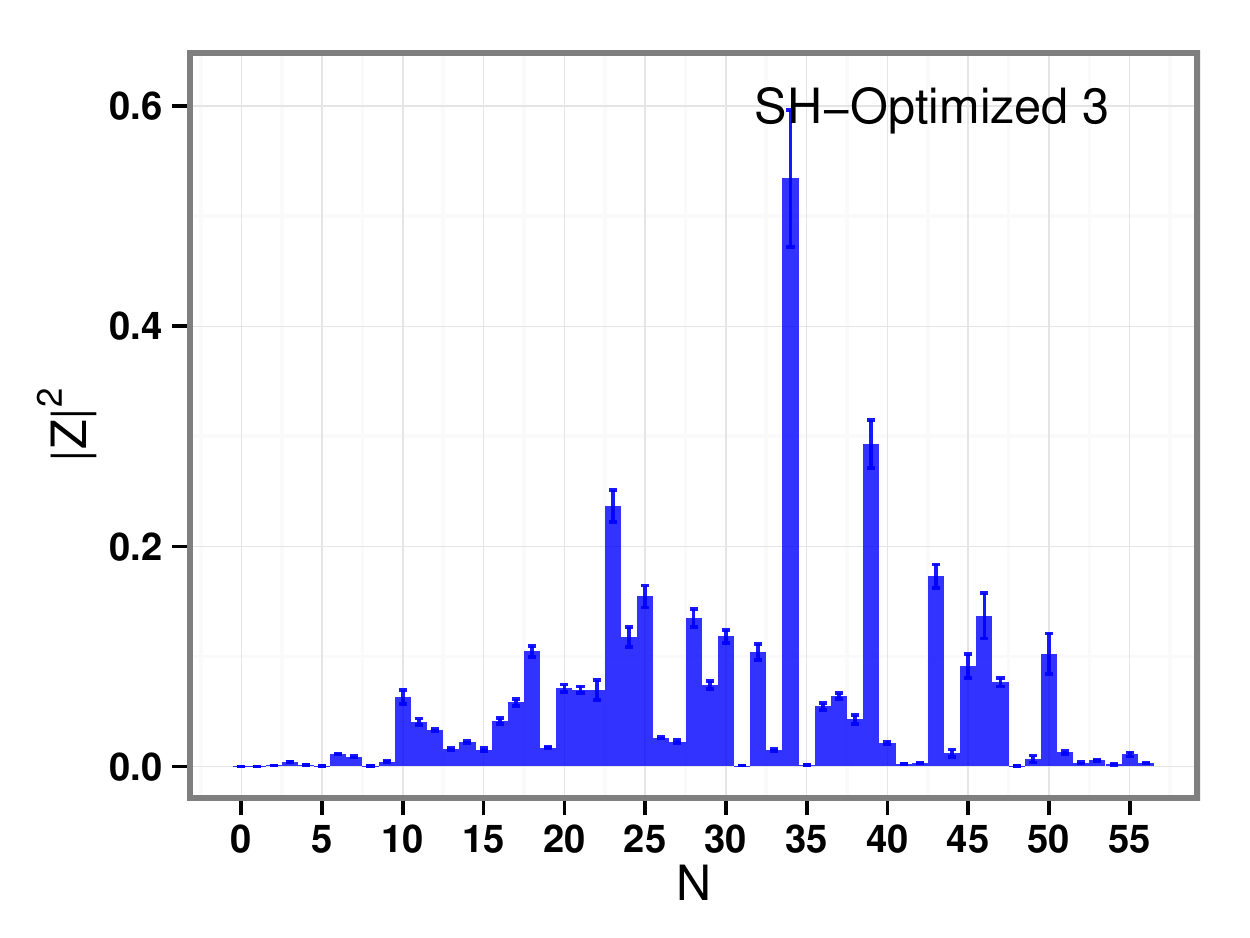}%
\\
\includegraphics[width=1.40in]{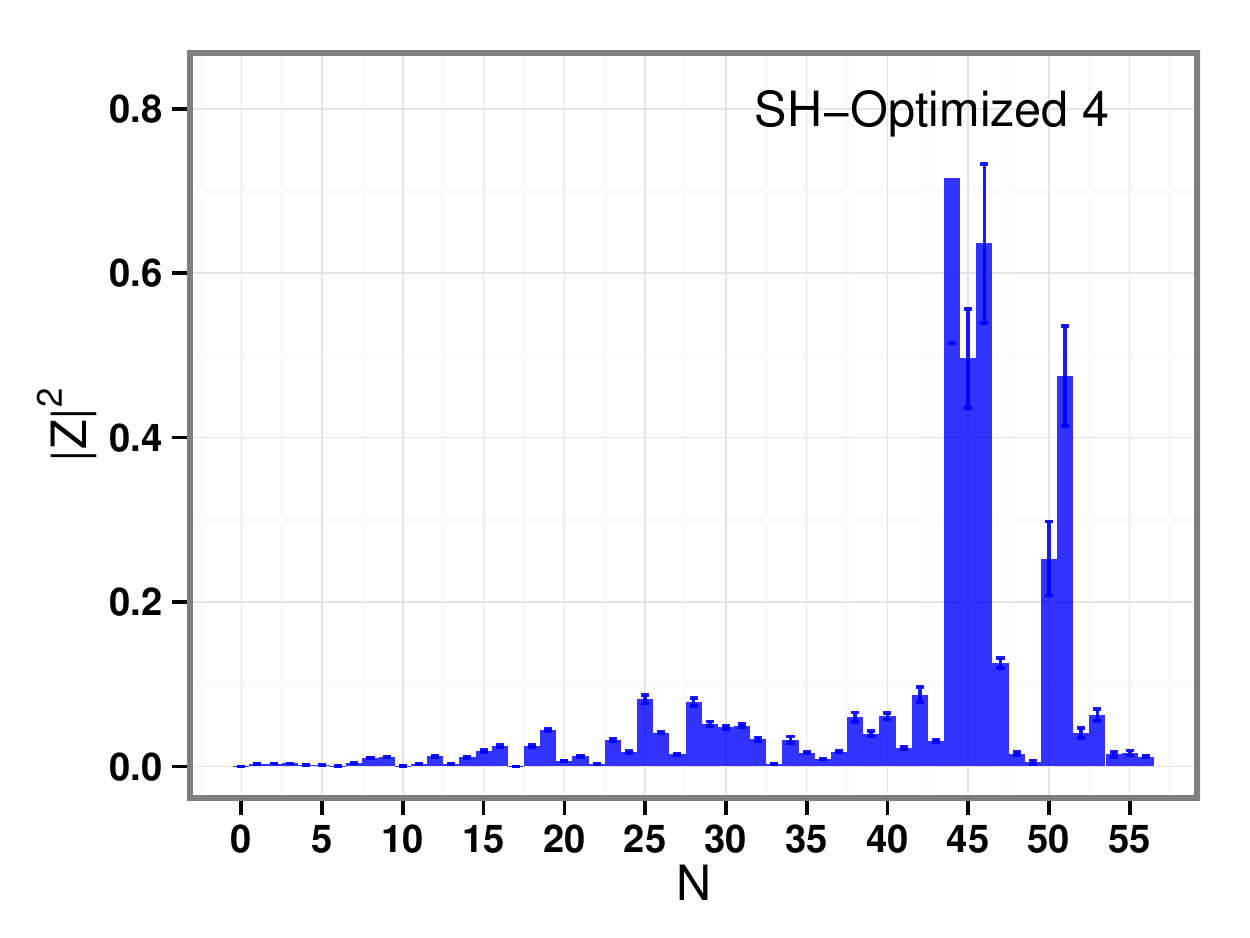}%
\includegraphics[width=1.40in]{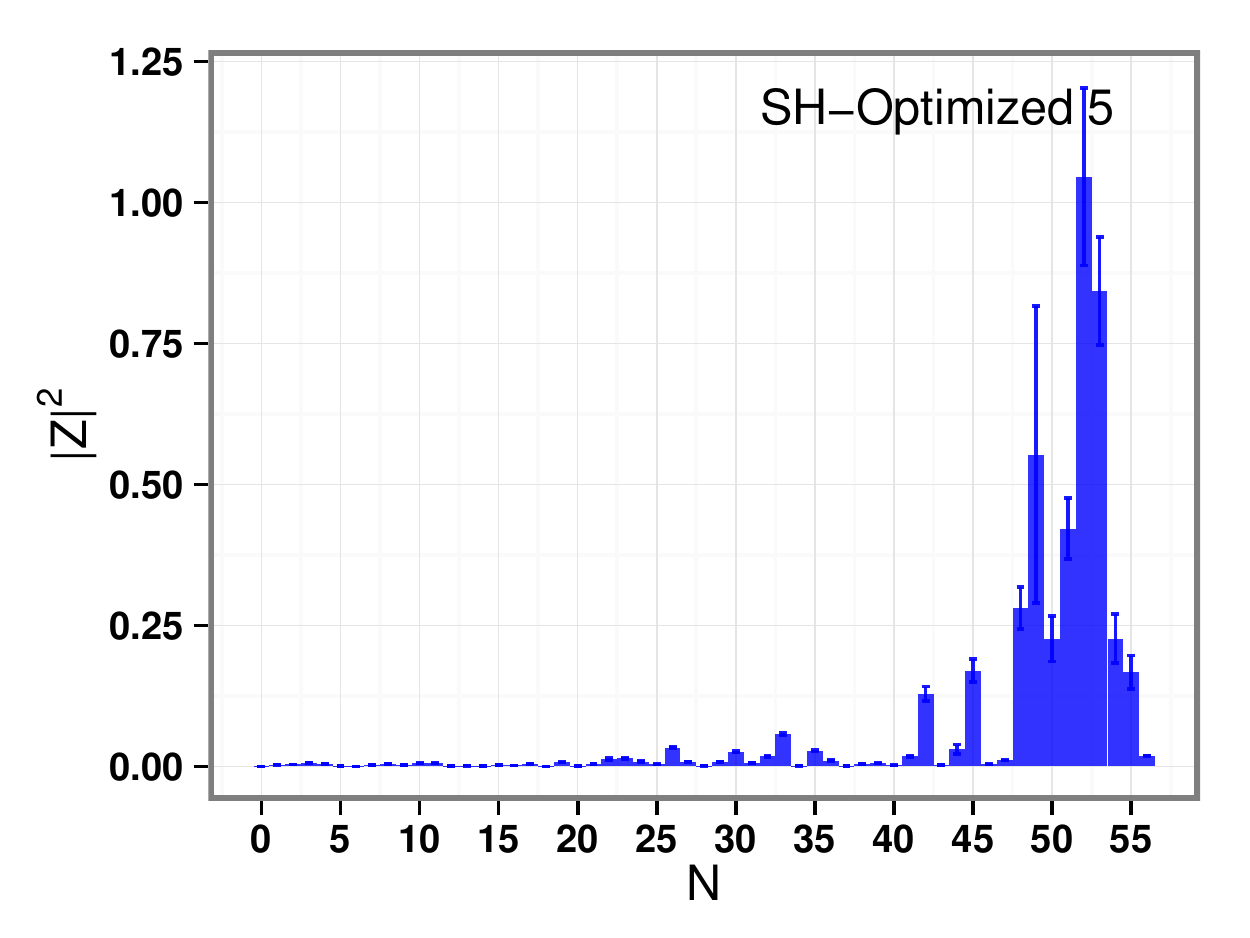}%
\includegraphics[width=1.40in]{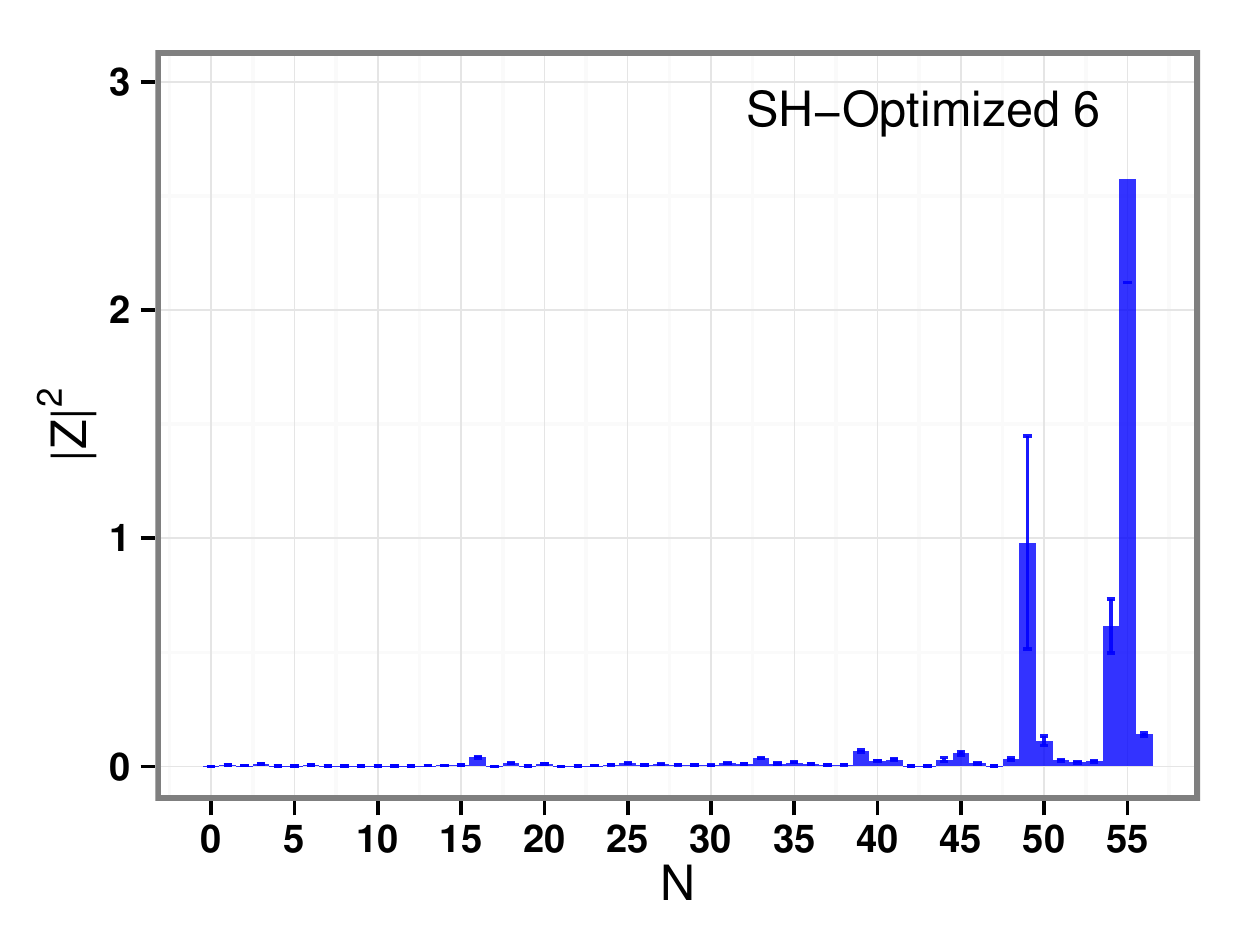}%
\includegraphics[width=1.40in]{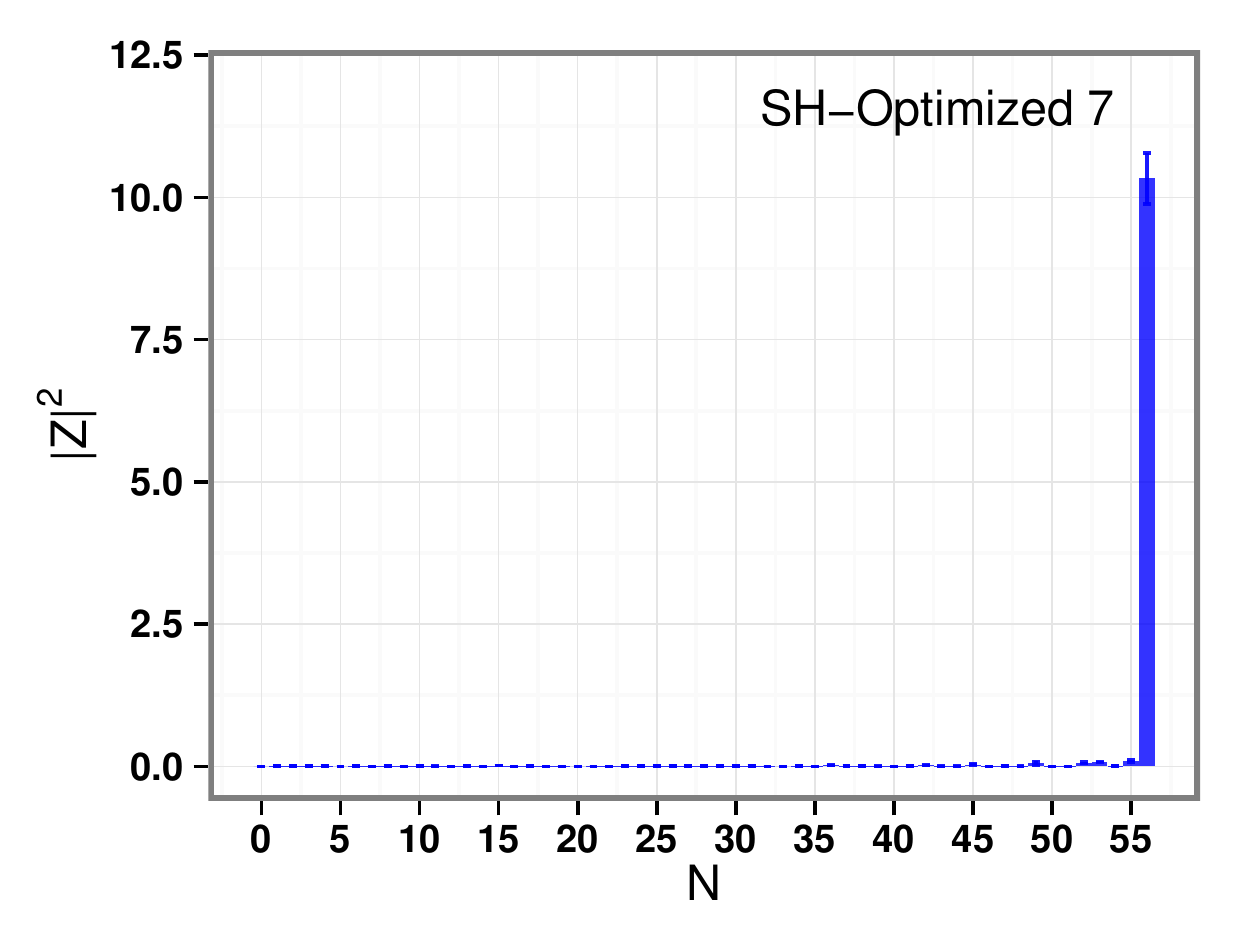}%
\end{center}
\vspace*{-7mm}
\caption[cap5]{
Overlaps $\vert \widetilde{Z}^{(n)}_j\vert^2$ of our ``optimized'' single-hadron
operator $\widetilde{O}_j$ against the eigenstates labelled by $n$.
The overall normalization is arbitrary in each plot.
\label{fig:Zopts}}
\end{figure}  

With such a large number of energies extracted, level identification 
becomes a key issue.  Level identification must
be inferred from the $Z$ overlaps of our probe operators.  
We first focus our efforts on identifying the levels that dominate
the finite-volume stationary states expected to evolve into 
the single-meson resonances in infinite volume.  We view such states
as ``resonance precursor states''.  To accomplish this, we utilize
``optimized'' single-hadron operators as our probes.  We first restrict
our attention to the $9\times 9$ correlator matrix involving only
the 9 chosen single-hadron operators.  We then perform an optimization
rotation to produce so-called ``optimized'' single-hadron (SH) operators
$\widetilde{O}_j$, which are linear combinations of the 9 original
operators.  We order these SH-optimized operators according
to their effective mass plateau values, then evaluate the overlaps 
$\widetilde{Z}_j^{(n)}$ for these SH-optimized operators using
our analysis of the full $58\times 58$ correlator matrix.  The
results are shown in Fig.~\ref{fig:Zopts}.

\begin{figure}[t]
\begin{center}
\includegraphics[width=1.8in]{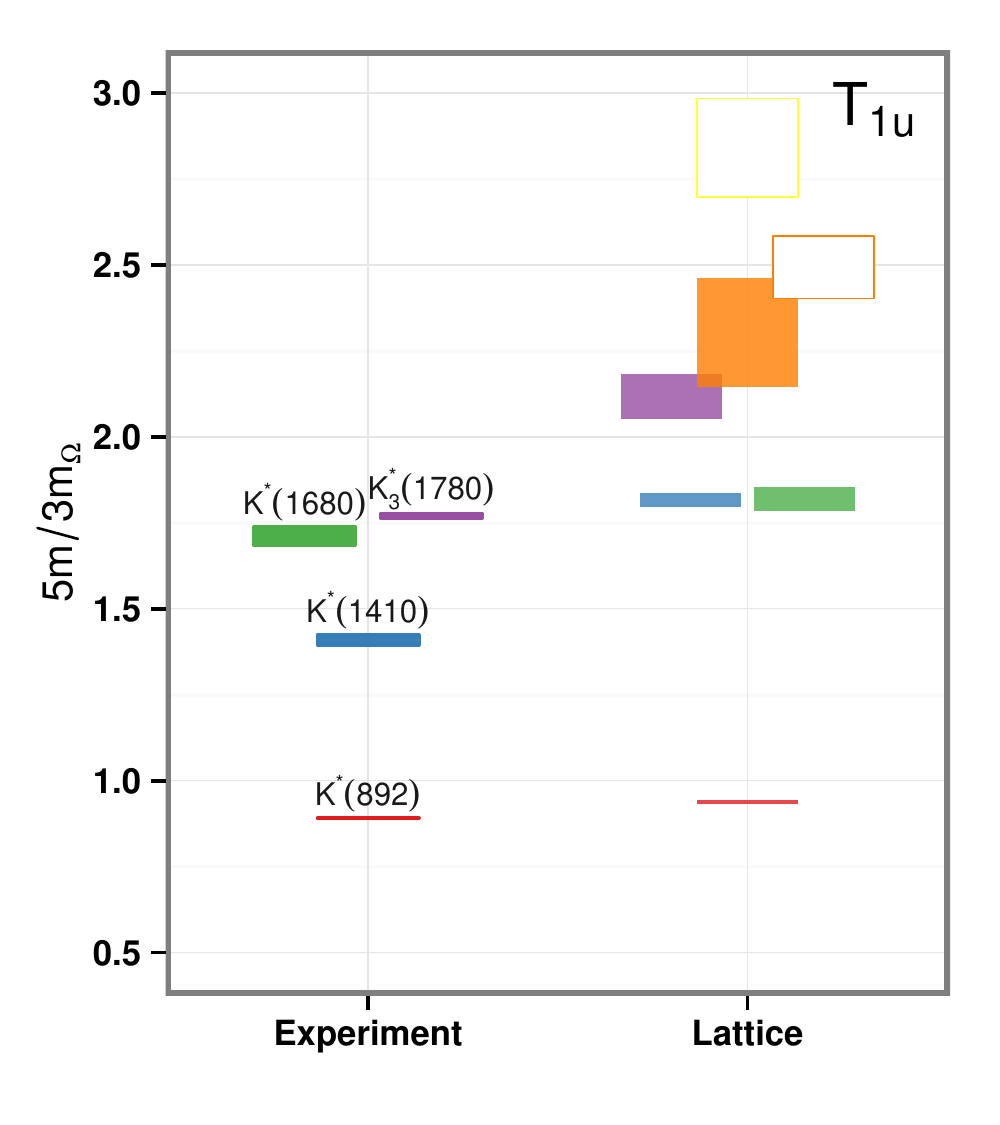}
\includegraphics[width=1.8in]{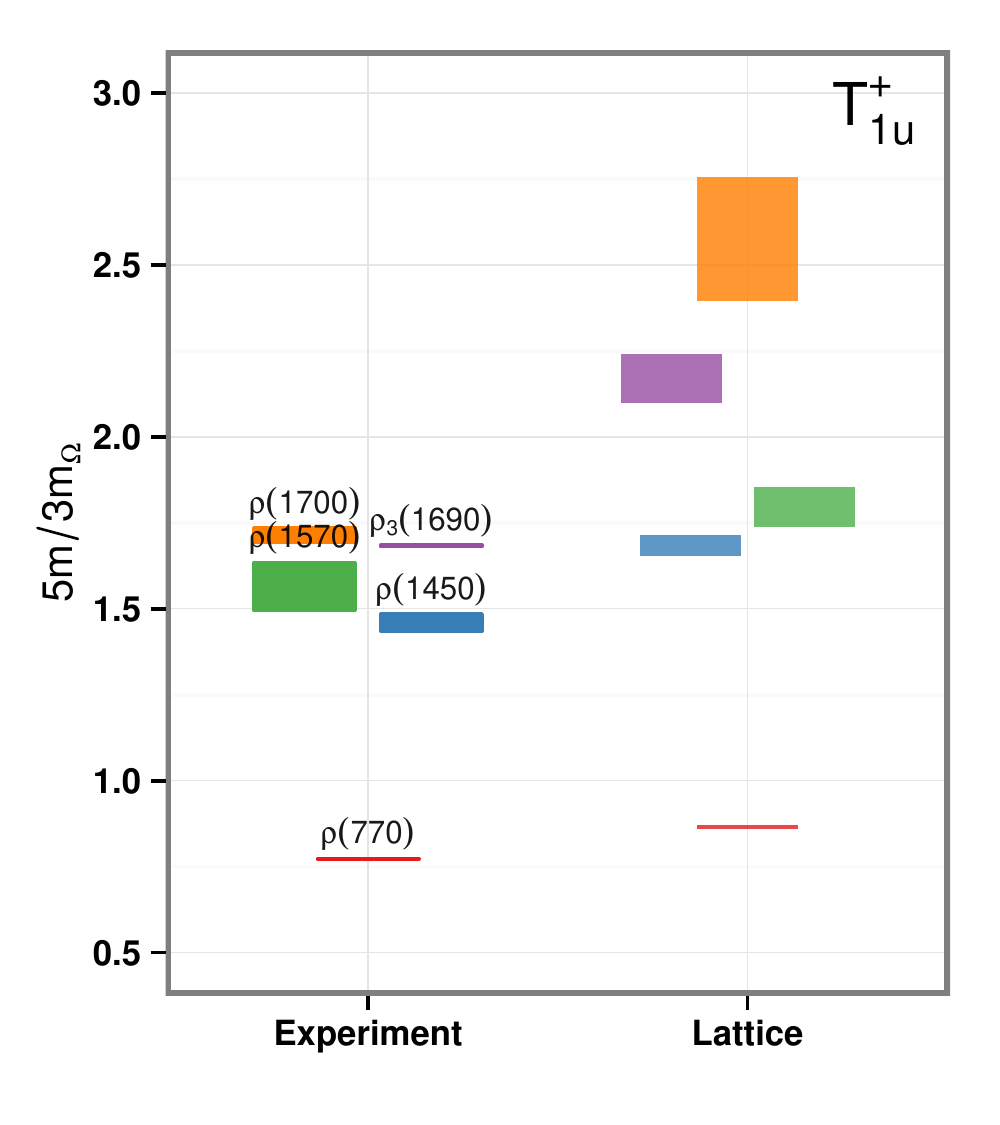}
\end{center}
\vspace*{-10mm}
\caption[cap4]{
(Left) Masses, as ratios with respect to 3/5 of the $\Omega$ baryon
mass $m_\Omega$, of the dominant
finite-volume isovector $T_{1u}$ stationary states expected to evolve into 
the single-meson resonances in infinite volume, computed using our
$58\times 58$ correlation matrix for the $(24^3\vert 390)$ ensemble.
The vertical thickness of each box indicates its statistical uncertainty.
The hollow boxes at the top show higher-lying states that we extract with 
less certainty due to the expected presence of lower-lying two-meson states
that have not been taken into account.
(Right) The analogous plot for the isovector zero-strangeness $T_{1u}^+$ 
channel, the superscript indicating $G$-parity.
\label{fig:boxplot}}
\end{figure}  

The first plot shows that the lowest-lying SH-optimized operator
produces level 0 and very little else.  Hence, we identify level 0
with the lowest-lying resonance precursor state, expected to be
the $K^\ast(892)$.  The second plot shows that this operator
produces mainly level 12.  Hence, we identify level 12 as the dominant
state that is the precursor of the first-excited resonance in this channel.
We summarize our single-hadron spectrum (the eigenstates dominated by the 
resonance precursor states) in Fig.~\ref{fig:boxplot}.  This figure shows the 
masses as a ratio of 3/5 of the $\Omega$ baryon mass.  Given that our pion 
mass is around 390~MeV and that our states are extracted in finite volume, 
precise agreement with experiment is certainly not expected.  These results
are compared to the analogous results in the isovector nonstrange
$T_{1u}^+$ channel from Ref.~\cite{lattice2013}.  Again, these
results are preliminary, and we mention that three and four meson states 
are not taken into account at all.

\section{Conclusion}
\label{sec:conclude}

In this talk, our progress in computing
the finite-volume stationary-state energies of QCD was described.
Preliminary results in the zero-momentum bosonic $I=\frac{1}{2},\ S=1,\ T_{1u}$ 
symmetry sector of QCD using a correlation matrix of 58 operators were presented.  
All needed Wick contractions were efficiently evaluated using 
the stochastic LapH method. Issues related to level identification 
were discussed.
This work was supported by the U.S.~NSF
under awards PHY-0510020, PHY-0653315, PHY-0704171, PHY-0969863, and
PHY-0970137, and through TeraGrid/XSEDE resources provided by 
TACC and NICS under grant numbers TG-PHY100027 and TG-MCA075017.

\end{document}